\documentclass[runningheads]{svmult}

\usepackage{makeidx}   
\usepackage{graphicx}  
\usepackage{subeqnar}  
\usepackage{multicol}  
\usepackage{physprbb}  
\makeindex             

\newcommand{\beq}{\begin{equation}}
\newcommand{\eeq}{\end{equation}}
\newcommand{\lesim}{\stackrel{<}{_\sim}}
\newcommand{\gesim}{\stackrel{>}{_\sim}}
\newcommand{\Gammash} {\Gamma_\mathrm{sh}}
\newcommand{\Gammarel}{\Gamma_\mathrm{rel}}
\newcommand{\betash}  {\beta_\mathrm{sh}}
\newcommand{\betarel} {\beta_\mathrm{rel}}
\newcommand{\mud}{\mu_{{\scriptscriptstyle\rightarrow}\rm d}}
\newcommand{\muu}{\mu_{{\scriptscriptstyle\rightarrow}\rm u}}
\newcommand{\thetad}{\theta_{{\scriptscriptstyle\rightarrow}\rm d}}
\newcommand{\thetau}{\theta_{{\scriptscriptstyle\rightarrow}\rm u}}

\begin{document}
\title*{Particle Acceleration at Relativistic Shocks}
\toctitle{Particle Acceleration at Relativistic Shocks}
%
%
\titlerunning{Particle Acceleration at Relativistic Shocks}
%
\author{Yves A. Gallant\inst{1,2}}
\authorrunning{Yves A. Gallant}
%
%
\institute{Osservatorio Astrofisico di Arcetri, Largo E. Fermi 5,
           50125 Firenze, Italy
\and       Astronomical Institute Utrecht, Postbus 80\,000,
           3508 TA Utrecht, Netherlands}

\maketitle              

\begin{abstract}
I review the current status of Fermi acceleration theory
at relativistic shocks.  I first discuss the relativistic
shock jump conditions, then describe the non-relativistic
Fermi mechanism and the differences introduced by relativistic
flows.  I present numerical calculations of the accelerated
particle spectrum, and examine the maximum energy attainable
by this process.  I briefly consider the minimum energy for
Fermi acceleration, and a possible electron pre-acceleration
mechanism.
\end{abstract}

\section{Introduction and Motivation}
\label{intro}
   A ubiquitous feature of astrophysical objects involving
relativistic flows, such as active galactic nuclei (AGNs),
gamma-ray bursts (GRBs) and Crab-like supernova remnants (SNRs),
is the presence of nonthermal, power-law emission spectra
(i.e. with flux density $F_\nu \propto \nu^{-\alpha}$, where
$\nu$ is the frequency and $\alpha$ the spectral index), in
particular in the radio and hard X-ray or gamma-ray domains.
This emission is believed to be produced by accelerated
particles having a corresponding power-law energy spectrum;
more specifically, in most of these objects the emission
is thought to be from accelerated electrons radiating via the
synchrotron or inverse Compton mechanisms (see Mastichiadis,
this volume).  The aim of the present review will be to
discuss the probable mechanism of this acceleration and
the spectra that may be expected theoretically.

   The most widely invoked mechanism for the acceleration
of particles to power-law spectra in non-relativistic contexts,
such as SNR blast waves or interplanetary shocks,
is \emph{Fermi acceleration}.  It seems likely that shocks
are responsible for particle acceleration in relativistic
flows as well, and this is indeed explicitly assumed in
models of GRBs and Crab-like SNRs.  It
is then natural to consider how the Fermi mechanism could
operate at relativistic shocks, and what the
resulting spectrum would be.  The focus of this
contribution will thus be the relativistic version of
Fermi shock acceleration.

   This review is organised as follows: in Sect.~\ref{relshocks},
I discuss the shock jump conditions at relativistic shocks,
emphasising the aspects relevant to particle acceleration;
this section is intended to be self-contained.  In Sect.~%
\ref{Fermi}, I describe the Fermi acceleration mechanism in
detail, first reviewing its main features in the context of
non-relativistic shocks, and then presenting the resulting spectrum
for ultra-relativistic and more moderately relativistic shocks.
In Sect.~\ref{maxmin}, I examine the acceleration time scale
and the maximum energy attainable by this mechanism, and
consider the minimum energy for Fermi acceleration of
electrons in an electron--ion shock, and a possible
pre-acceleration mechanism.

\section{Relativistic Shocks}
\label{relshocks}
The properties of shocks most important for Fermi-type particle 
acceleration are the velocities of the shock relative to the upstream
and downstream frames.  These are obtained through the \emph{shock jump
conditions}.

\subsection{Relativistic Shock Jump Conditions}
The shock jump conditions are derived from the laws of conservation
of particle number, energy, and momentum.  For relativistic fluids,
these are, in order:
\begin{eqnarray}
\label{numcons}
\Gamma_1 \beta_1 n_1 & = & \Gamma_2 \beta_2 n_2  \; ,
\\
\label{encons}
\Gamma_1^2 \beta_1 (\varepsilon_1 + p_1) & = &
\Gamma_2^2 \beta_2 (\varepsilon_2 + p_2)  \; ,
\\
\label{momcons}
\Gamma_1^2 \beta_1^2 (\varepsilon_1 + p_1) + p_1 & = &
\Gamma_2^2 \beta_2^2 (\varepsilon_2 + p_2) + p_2 \; ,
\end{eqnarray}
where subscripts 1 and 2 respectively refer to the upstream and
downstream regions, $n$, $\varepsilon$ and $p$ are the fluid number
density, energy density and pressure, all measured in the local fluid
rest frame, $\beta$ is the fluid velocity in units of the speed
of light $c$, and $\Gamma$ the corresponding Lorentz factor. The
fluid velocities are measured in the \emph{shock frame}, where
the shock is stationary and both velocity vectors lie along the
shock normal.  Equivalently, $\beta_1$ and $\beta_2$ may be viewed
as the shock velocity with respect to the upstream and downstream
fluids.

   For simplicity, I restrict my attention in this review to
unmagnetised shocks.  The shock jump conditions for relativistic
magneto-hydrodynamics (MHD) were reviewed by Kirk and Duffy~%
\cite{KirDuf99}.  They can yield shock jump conditions which
differ significantly from those derived here when the magnetisation
parameter,
\beq
\label{sigma}
\sigma \equiv \frac{B_1^2}{4\pi (\varepsilon_1 + p_1)} \; ,
\eeq
where $B$ is the magnetic field measured in the local fluid rest
frame, is not negligibly smaller than unity~\cite{KenCor84,Kiretc00}.

\subsection{Ultra-Relativistic Shocks}
\label{URshocks}
Much of the discussion of particle acceleration below will be
specialised to ultra-relativistic shocks, i.e.\ those in which
the shock Lorentz factor $\Gammash \equiv \Gamma_1 \gg 1$, and
$\beta_1 \approx 1$.  In that case the pressure term $p_1$ may
be neglected in (\ref{momcons}) relative to the first term.
The particles downstream of such a shock must be heated to
highly relativistic temperatures; assuming they obey the
ultra-relativistic gas equation of state, $\varepsilon_2 = 3 p_2$,
one may then solve (\ref{encons}) and (\ref{momcons}) to obtain
the \emph{ultra-relativistic shock jump conditions}, yielding
the downstream velocity
\beq
\beta_2 \approx \frac{1}{3} \; .
\eeq
Also of interest is the relative velocity $\betarel$ of the
upstream and downstream fluids.  Using the relativistic
velocity addition formula,
\beq
\label{betarel}
\betarel = \frac{\beta_1 - \beta_2}{1 - \beta_1 \beta_2} \; ,
\eeq
the associated Lorentz factor in the ultra-relativistic
limit is $\Gammarel \approx \Gammash / \sqrt{2}$.

   It should be noted that these shock jump conditions are
independent of the upstream equation of state, and depend only
on the downstream gas being ultra-relativistically hot.  It
will be seen below that particle acceleration in the
ultra-relativistic shock regime thus mirrors some of the
simplicity of the non-relativistic, strong shock regime, due
to the existence of this single, well-defined asymptotic value
(for weakly magnetised shocks) of the shock velocity ratio.

\subsection{Moderately Relativistic Shocks}
\label{MRshocks}
For more general values of the shock Lorentz factor $\Gammash$,
the shock jump conditions depend on the equation of state and
the temperature of the upstream gas.  For illustration, the
two opposite extremes of an ultra-relativistically hot and a
cold gas upstream will be examined.

\subsection*{Shocks in an Ultra-Relativistic Gas}
I first assume that the gas upstream of the shock already has
a highly relativistic temperature; this might be the case for
an internal shock in a GRB fireball, for instance, if it
propagates in a medium already heated by previous shell
collisions (see e.g.\ Sari \& Galama, this volume).
In this case, both the upstream and the downstream media can
be assumed to follow the ultra-relativistic equation of state,
$\varepsilon = 3 p$.  Equations (\ref{encons}) and (\ref{momcons})
are then readily solved to yield the jump condition for shocks
propagating in an ultra-relativistic gas:
\beq
\label{URgas}
\beta_1 \beta_2 = \frac{1}{3} \; .
\eeq

   This relation holds for shocks of any strength, provided
only that the upstream gas is ultra-relativistic.  The only
requirement for the existence such of a shock solution is
that the upstream flow velocity be larger than the sound
speed in the upstream gas, which is $c / \sqrt{3}$ for an
ultra-relativistic gas.

\subsection*{Strong Shocks and the Synge Equation of State}
I now consider the case of a \emph{strong shock}, i.e. one
in which the thermal energy upstream is negligible with respect
to the bulk flow kinetic energy, so that one may neglect the
upstream pressure $p_1$ and write $\varepsilon_1 \approx n_1
m c^2$ in (\ref{encons}) and (\ref{momcons}), where $m$ is
the mass of individual gas particles, assumed for simplicity
to belong to a single species.
For the downstream equation of state, I use that of an ideal
gas of arbitrary temperature, as given by Synge \cite{Syn57}:
\beq
\label{SyngeEOS}
\varepsilon_2 + p_2 = n_2 m c^2
                      G\left( \frac{m c^2}{T_2} \right) \; .
\eeq
Here $T_2$ is the downstream gas temperature, and the 
function $G(\xi)$ is defined in terms of modified Bessel
functions of the first kind,
$ G (\xi) \equiv {K_3 (\xi)} / {K_2 (\xi)} $,
and has the asymptotic expansions:
\begin{eqnarray}
G \left( \frac{m c^2}{T} \right)  & = & 1 + \frac{5}{2}
\frac{T}{m c^2} + \mathcal{O}\left( \frac{T}{m c^2} \right)^2
\; , \qquad \quad  T \ll m c^2 \; , \\
G \left( \frac{m c^2}{T} \right) & = & \frac{4T}{m c^2} +
\frac{m c^2}{2 T} + \mathcal{O}\left( \frac{m c^2}{T}\right)^3
\; , \qquad \;     T \gg m c^2 \; .
\end{eqnarray}
Using the fact that the gas always obeys the ideal gas law
$p = n T$, it is readily seen that these two asymptotes
correspond to the familiar equations of state for
non-relativistic and ultra-relativistic ideal gases,
namely $\varepsilon = n m c^2 + 3 p / 2$ and
$\varepsilon = 3 p$, respectively.

\begin{figure}[t]
\begin{center}
\includegraphics[height=7cm]{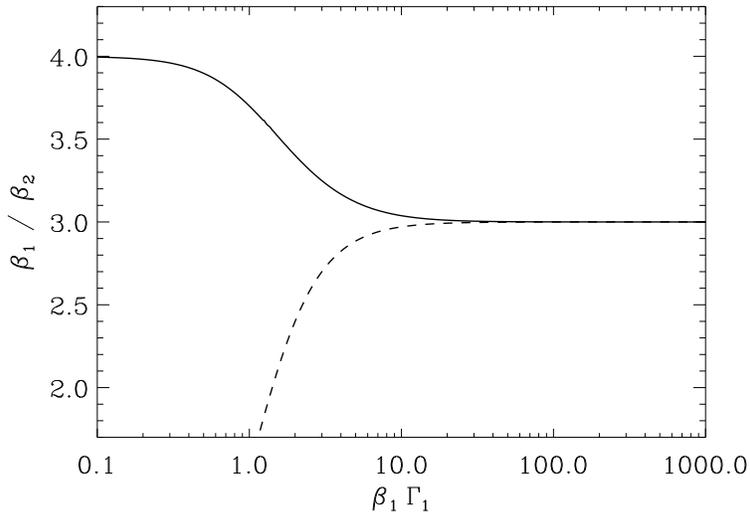}
\end{center}
\caption{\label{velrat}
The shock velocity ratio as a function of the upstream
four-velocity, for the two extreme cases of a cold gas
upstream obeying the Synge equation of state downstream
({\it solid line\/}), and of an ultra-relativistically
hot gas upstream ({\it dashed line\/})}
\end{figure}

   With the above assumptions, the shock jump conditions 
(\ref{numcons}--\ref{momcons}) may be solved by first using
(\ref{numcons}) to rewrite (\ref{encons}) and (\ref{momcons})
in terms of the normalised quantities
\begin{eqnarray}
\label{ebar}
\bar{\varepsilon}_2 & \equiv & \frac{\varepsilon_2}{n_2 m c^2}
 = G(\xi) - \frac{1}{\xi} \; , \\
\bar{p}_2 & \equiv & \frac{p_2}{n_2 m c^2} = \frac{1}{\xi} \; ,
\end{eqnarray}
where the \emph{reciprocal temperature} of the downstream gas
has been defined as $\xi \equiv m c^2 / T_2$.  The two resulting
equations may be solved to yield $\Gamma_1$ and $\Gamma_2$
in terms of $\xi$ \cite{KirDuf99}:
\begin{eqnarray}
\label{Gamma2}
\Gamma_2^2 & = & \frac{\bar{\varepsilon}_2^2 - 1}%
                 {\bar{\varepsilon}_2^2 - \bar{p}_2^2 - 1}
\; , \\
\label{Gamma1}
\Gamma_1   & = & (\bar{\varepsilon}_2 + \bar{p}_2) \Gamma_2
\; .
\end{eqnarray}
If one prefers to use $\Gamma_1$ rather than $T_2$ as the
independent variable, the analytical equation (\ref{Gamma1}),
substituting the definitions (\ref{ebar}--\ref{Gamma2}), may
be inverted numerically.
The shock velocity ratio $\beta_1/\beta_2$ resulting from
(\ref{Gamma2}--\ref{Gamma1}) is shown in Fig.~\ref{velrat}
as a function of the upstream four-velocity $\beta_1 \Gamma_1$,
along with the corresponding result for the ultra-relativistic
upstream gas case (\ref{URgas}).

\subsubsection*{Electron--Ion Plasmas:}
   The shock jump conditions derived above, based on the
equation of state (\ref{SyngeEOS}), are strictly speaking
only valid for a gas composed of particles of a single mass $m$,
as would be the case for instance in an electron--positron plasma.
Equation (\ref{SyngeEOS}) is readily generalised to a gas
composed of species of different masses \cite{KirDuf99,Syn57},
assuming they are in thermal equilibrium.
However, simulations of relativistic, perpendicular shocks in
electron--ion plasmas \cite{Hosetc92} show that the species do
not in fact achieve thermal equilibrium immediately behind the
shock, but instead have distinct temperatures corresponding
to the thermalisation of their respective upstream bulk kinetic
energy.  The strong shock jump condition illustrated in
Fig.~\ref{velrat} then also applies in this case, as I now
demonstrate.

   As remarked in \cite{KirDuf99}, for a strong shock one may
derive from the shock jump conditions (\ref{numcons}--\ref{momcons})
the relation
\beq
\varepsilon_2 = \Gammarel \rho_2 c^2 \; ,
\eeq
where $\rho_2$ is the total downstream rest mass density, and
$\Gammarel$ the Lorentz factor corresponding to the relative
velocity (\ref{betarel}).  This equation shows that the downstream
energy density per unit mass is simply the upstream bulk flow
energy of the particles as seen from the downstream frame.
If this relation holds separately for each species, as is the
case initially in the electron--ion shock simulations mentioned
above, one has for each species, using (\ref{ebar}):
\beq
G(\xi) - \frac{1}{\xi} = \Gammarel \; .
\eeq
Thus although the temperatures of the species will in general
be different, their normalised (reciprocal) temperatures $\xi$
will be the same, and the shock jump conditions obtained above
for a single particle mass will hold in this case also.

   Energy exchange between the electrons and the ions does
take place downstream of the shock in the above-mentioned
simulations, but appears to result in a power-law tail of
the electron energy distribution rather than simple heating,
as discussed further in Sect.~\ref{preacc}.  One could
envision shock jump conditions taking into account this 
phenomenon in the downstream equation of state; however,
an equally important component of more realistic shock
jump conditions is the energy and momentum carried away
by the strong electromagnetic precursor emitted by the
shock front \cite{Galetc92}.  As neither of these two
phenomena can be predicted quantitatively at present, it
seems premature to attempt to obtain more accurate shock
jump conditions for electron--ion plasmas than that shown
in Fig.~\ref{velrat} for a simple Synge equation of state.

\section{Fermi Acceleration and the Spectral Index}
\label{Fermi}
In this section, I first review the basic ideas of the Fermi
acceleration mechanism in the context of non-relativistic
shocks, then discuss in some detail its application to the
opposite extreme of ultra-relativistic shocks and the spectral
index resulting in that case, before addressing the more
involved intermediate case of moderately relativistic shocks.

\subsection{Non-Relativistic Shocks} 
While the essential concepts of the acceleration mechanism that
bears his name date back to Fermi \cite{Fer49}, their application
to shock acceleration was first proposed in 1977--78 in four
independent papers \cite{AxfLeeSka78,Bel78a,BlaOst78,Kry77};
of these, I will follow most closely below the treatment given
by Bell \cite{Bel78a}.

\begin{figure}[t]
\begin{center}
\includegraphics[height=7cm]{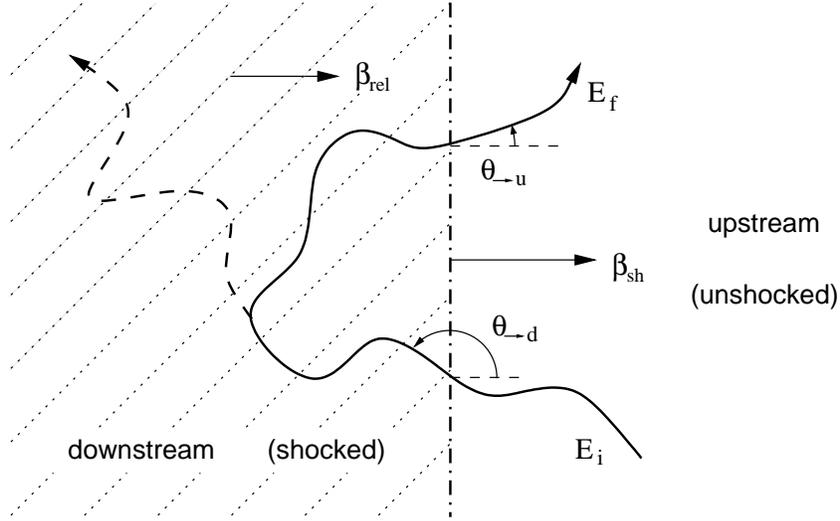}
\end{center}
\caption{\label{accschem}
Schematic representation of one cycle of shock acceleration:
a shock propagates with velocity $\beta_\mathrm{sh}$ into
the undisturbed (upstream) medium to the right; the velocity
of the shocked (downstream) medium relative to the upstream
one is $\betarel$.  A relativistic particle diffuses through
the media on both sides of the shock, crossing and re-crossing
it at incidence angles $\thetad$ and $\thetau$, with initial
and final energies $E_\mathrm{i}$ and $E_\mathrm{f}$}
\end{figure}

   The acceleration scenario is illustrated schematically
in Fig.~\ref{accschem}: a high-energy particle, assumed for
simplicity to be already relativistic, diffuses through the
medium on either side of the shock
by scattering on magnetic irregularities.  These may
be, for instance, Alfv\'en waves self-consistently excited
by the diffusing high-energy particles \cite{Bel78b}.
Assuming the local Alfv\'en velocity is much smaller than
the shock velocity, to lowest order the magnetic scattering
centres may be considered at rest with respect to the fluid,
so that the scattering events do not change the particle
energy in the local fluid rest frame.

   Consider, then, a particle diffusing upstream or downstream
while preserving its energy in the corresponding rest frame.
A particle initially having energy $E_\mathrm{i}$ in the
upstream medium will eventually cross the shock, its velocity
upon crossing making an angle $\thetad$ with the shock normal
(see Fig.~\ref{accschem}).  Its energy measured in the
downstream frame, $E_\mathrm{i}'$, will then be given by
the appropriate Lorentz transformation, and preserved while
the particle is downstream.  If it re-crosses the shock into
the upstream medium, this time at an angle $\thetau'$, its
final energy upstream, $E_\mathrm{f}$, will be given by the
combination of the two Lorentz transformations:
\beq
\label{Egain-gen}
\frac{E_\mathrm{f}}{E_\mathrm{i}} = \Gammarel^2
( 1 - \betarel \mud ) ( 1 + \betarel \muu' ) \; ,
\eeq
where $\betarel$ and $\Gammarel$ are the relative velocity
of the upstream and downstream media and the corresponding
Lorentz factor, and I have introduced the notation $\mu$
for $\cos \theta$.  Here and in what follows primed and
unprimed quantities are respectively measured in the downstream
and upstream rest frames.  The only approximation made in deriving
(\ref{Egain-gen}) is that that the particle is highly relativistic,
so that the rest mass contribution to its energy may be neglected.

   For non-relativistic shocks, $\betarel \ll 1$, the angular
distribution of these scattered particles crossing the shock may be
approximated as isotropic, so that the flux-weighted averages
of the direction angle cosines, over the relevant ranges
$-\pi/2 \leq \thetad \lesim 0$ and $0 \lesim \thetau' \leq \pi/2$,
are respectively $\langle \mud \rangle \approx -2/3$ and $\langle
\muu' \rangle \approx 2/3$.  The average energy gain (\ref{Egain-gen})
per shock crossing cycle then reduces to
\beq
\label{Egain-NR}
\left\langle \frac{E_\mathrm{f}}{E_\mathrm{i}} \right\rangle
\approx 1 + \betarel
\left( \langle \muu' \rangle - \langle \mud \rangle \right)
\approx 1 + \frac{4}{3} \betarel \; .
\eeq
Fermi shock acceleration is sometimes referred to as the
\emph{Fermi I} mechanism because this energy gain is of first
order in the velocity $\betarel$.

   While a particle upstream will always eventually cross the
shock, at least in the simple case considered here of an infinite,
plane-parallel shock, once downstream the particle has a certain
probability of being advected away and never re-crossing the shock.
This \emph{escape probability} may be evaluated from the ratio of
the average fluxes of particles escaping far downstream and crossing
the shock.  For a downstream flow velocity $\beta_2$, again assuming
isotropy of the relativistic particle distribution, it is given by
$P_\mathrm{esc} = 4 \beta_2$.

   The combination of the energy gain factor (\ref{Egain-NR}) and
escape probability $P_\mathrm{esc}$ leads to a power-law spectrum
in the accelerated particle energy, with a spectral index depending
solely on the shock jump conditions:
\beq
\frac{\D N}{\D E} \propto E^{-(r+2)/(r-1)} \; ,
\eeq
where $r \equiv \beta_1 / \beta_2$ is the shock \emph{velocity
ratio}.  For a strong, non-relativistic shock in a monatomic gas,
$r=4$, leading to a spectral index $(r+2)/(r-1) = 2$ for the
accelerated particles.
In reality, spectral indices somewhat steeper than this value
are often observed; this difference may be due to the pressure
of the accelerated particles modifying the shock structure
(e.g. \cite{BerEll99} and references therein).

\subsection{Ultra-Relativistic Shocks}
   I now turn my attention to ultra-relativistic 
shocks, i.e. those for which $\Gammash \gg 1$ so that the shock
jump conditions derived in Sect.~\ref{URshocks} apply.

\subsection*{Energy Gain and Upstream Particle Dynamics}
\label{E-updyn}
   For a downstream particle to cross the shock into the
upstream medium,
it must have $1 \geq \muu' > \betash' = \frac{1}{3}$, so that
the factor $(1 + \betarel \muu')$ in (\ref{Egain-gen}) is always
of order unity.  If $\mud$ is approximately isotropically
distributed, as might be the case for a population of relativistic
particles already present in the undisturbed upstream medium,
the factor $(1 - \betarel \mud)$ is in general also of order unity.
Thus in the first shock crossing cycle, a large \emph{initial
boost} in energy can be achieved, $E_\mathrm{i} / E_\mathrm{f}
\sim \Gammarel^2$ as envisioned in \cite{Vie95}.

   For all subsequent shock crossing cycles, however, the
distribution of $\mud$ will be highly anisotropic; this is an
essential difference between non-relativistic and relativistic
Fermi shock acceleration \cite{KirSch87,Pea81}.  For an
ultra-relativistic particle with Lorentz factor $\gamma \gg
\Gammash$, the kinematic condition to cross the shock into the
upstream medium reduces to $\thetau < 1/\Gammash$.  As
shown in \cite{GalAch99}, for realistic deflection processes
upstream the particle cannot be deflected very far beyond
this `loss cone' before the shock overtakes it, so that
$\thetad \sim 1/\Gammash$ as well.  In this case the energy
gain factor reduces to
\beq
\label{Egain-UR}
\frac{E_\mathrm{f}'}{E_\mathrm{i}'} \approx
\frac{2 + (\Gammash \thetad)^2}{2 + (\Gammash \thetau)^2}
\approx \frac{1 + \muu'}{1 + \mud'} \; ,
\eeq
where the shock crossing cycle is now considered from
downstream to upstream and back.

   The range of possible energy gain factors can be assessed
by considering two opposite extremes for the upstream particle
dynamics: deflection by a regular magnetic field and scattering
by small-scale magnetic fluctuations.  In terms of the correlation
length $\ell$ of the magnetic field, these two regimes respectively
correspond to $R_\mathrm{L}/\Gammash \ll \ell$ and
$R_\mathrm{L}/\Gammash \gg \ell$, where $R_\mathrm{L}$ is the Larmor
radius of the particle.  For regular deflection, it can be shown
that for ingress angles $0 \leq \Gammash \thetau < 1$, the egress
angle satisfies
\beq
1 < \Gammash \thetad \leq 2
\qquad \qquad \Longleftrightarrow \qquad \qquad
\frac{1}{3} > \mud' \geq -\frac{1}{3} \; ,
\eeq
while for direction-angle scattering, the direction angle at the
average shock recrossing time satisfies
\beq
\langle \thetad^2 \rangle \approx \frac{2}{\Gammash^2} - \thetau^2 \; .
\eeq
In both cases, it may be seen that the typical energy gain
$\Delta E' \equiv E_\mathrm{f}' - E_\mathrm{i}'$ is thus of the
order of $E_\mathrm{i}'$ itself \cite{GalAch99}.

\subsection*{Numerical Calculation of the Angular Distribution}
   As was seen in the case of non-relativistic shock acceleration,
the power-law index of the accelerated particle distribution
depends on the average energy gain per shock crossing and the
escape probability.  For relativistic shocks, both of these are
strongly dependent on the angular distribution of particles
crossing the shock, which as suggested in the previous section
is in general highly anisotropic.  Thus the quasi-isotropic
approximations used in the non-relativistic case do not apply
here, and the distribution of the shock crossing angles $\muu$
and $\mud$ has been evaluated numerically.

   For simplicity, I will focus in this section on the case
where both the upstream and downstream particle dynamics are
dominated by scattering of the particle momentum direction.
In other words, it is assumed that magnetic fluctuations
dominate over the regular magnetic field in determining the
particle transport, at least along the shock normal direction
which is of interest here.  This is a highly plausible assumption
downstream, where shock-generated turbulence is likely to
give rise to disordered magnetic fields significantly stronger
than the shock-compressed upstream field, as is assumed in
relativistic fireball models of gamma-ray bursts and their
afterglows (see Sari \& Galama, this volume).

\begin{figure}[t]
\begin{center}
\includegraphics[height=7cm]{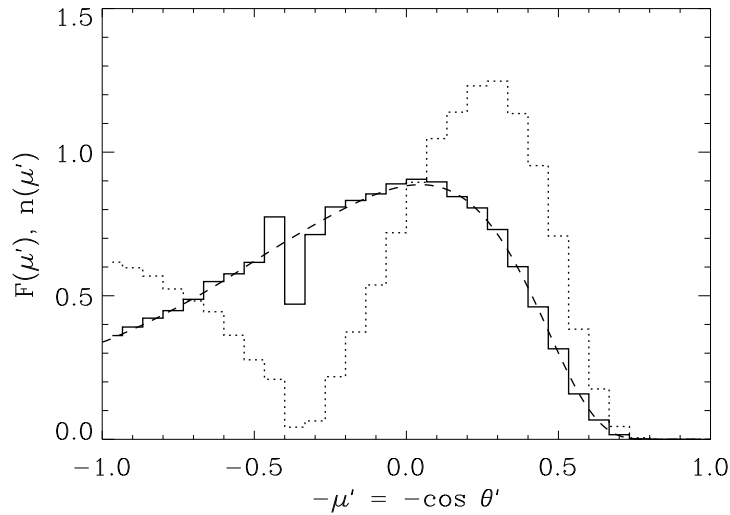}
\end{center}
\caption{\label{angdist}
Asymptotic downstream angular distribution of the particles
crossing the shock, showing both the flux $F(\mu')$ ({\it
dotted line\/}) and density $n(\mu')$ ({\it solid line})
obtained by Monte-Carlo simulations, along with the density
obtained by the eigenfunction method ({\it dashed line}).
All distributions are normalised to unity
}
\end{figure}

   One method of computing the accelerated particle distribution
is through numerical simulations, and I illustrate below the
results of such a calculation, after summarising the algorithm:
since the nature of the particle transport upstream and
downstream is by assumption independent of particle energy,
it is computationally more efficient to decouple the dynamical
problem from the energy gains.  A numerical approximation to the
function $f_\mathrm{d} (\muu' ; \mud')$, the distribution of
downstream egress angles $\muu'$ for a given ingress angle, is
thus first constructed by Monte-Carlo simulation of the downstream
scattering process for a grid of $\mud'$ values.  The upstream
dynamics are represented by a similarly obtained upstream egress
angle distribution $f_\mathrm{u} (\mud ; \muu)$, and both
distributions, along with the energy gain formula (\ref{Egain-UR}),
are subsequently used in a Monte-Carlo calculation of the
steady-state flux of accelerated particles crossing the shock.

   The results of such a calculation were summarised in
\cite{Galetc00}.  The influence of the highly anisotropic
injected particle distribution was seen to disappear at a
little more than a decade above the downstream injection
energy, at which point the self-consistent angular
distribution was established with a power-law in energy,
$F(E',\mu') \propto F(\mu') E'^{-p}$.  Here $F$ represents
the steady-state flux of accelerated particles crossing
the shock, per unit energy $E'$ and direction angle
cosine $\mu'$.  The asymptotic angular distribution
obtained with this simulation method is displayed in
Fig.~\ref{angdist}, which shows both $F(\mu')$ and the
corresponding density distribution, $n(\mu') \propto
F(\mu') / (\mu' - \betash')$.  The latter is compared
with the distribution obtained with the very different
semi-analytical eigenfunction method of Kirk et al.\
\cite{Kiretc00}, showing excellent agreement between
the two methods.

\subsection*{Spectral Index and Comparison with Observations}
   In the case considered above of isotropic scattering
upstream and downstream of the shock due to a strongly
turbulent magnetic field, a value of the spectral index
$p = 2.23 \pm 0.01$ is found by both Monte-Carlo
simulations \cite{Achetc01,Galetc00} and the
semi-analytical eigenfunction method \cite{Kiretc00}.
For the opposite extreme in upstream dynamics of
deflection by a regular magnetic field (see
above), still assuming isotropic scattering
downstream, Monte-Carlo simulations yield $p = 2.30$
\cite{Achetc01}.  With particle transport including
both a regular field and magnetic fluctuations,
Bednarz and Ostrowski \cite{BedOst98} obtained in
Monte-Carlo simulations values of $p \approx 2.2$
in the limit $\Gammash \gg 1$.  A spectral index
in the range $p = 2.2$--$2.3$ is thus a general
feature of Fermi acceleration at
(weakly magnetised) ultra-relativistic
shocks, at least in the test-particle approximation
used in all the above studies.

   Spectral index values deduced from observations of
astrophysical systems thought to involve ultra-relativistic
shocks are consistent with these theoretical expectations.
Early modelling of gamma-ray burst afterglow observations
suggested $p = 2.3 \pm 0.1$ \cite{Wax97}, and detailed
analysis of the GRB 970508 afterglow spectrum yielded
$p = 2.2$ \cite{Galetc98}.  While an equally detailed
multi-wavelength spectral analysis has not been published
for other afterglows, a value of $p \approx 2.2$ seems
compatible with most \cite{FraGal00}.
In Crab-like supernova remnants, the inferred spectral
indices are similar: the best-fit model for the Crab
Nebula spectrum corresponds to $p$ in the range 2.2--2.3
\cite{KenCor84}.
The very good agreement between theory and observation
is all the more remarkable given that in non-relativistic
shocks, as mentioned above, the observed particle
spectra often differ from the predictions of the simple
test-particle theory.

\subsection{Moderately Relativistic Shocks}
\begin{figure}[b]
\begin{center}
\includegraphics[height=7cm]{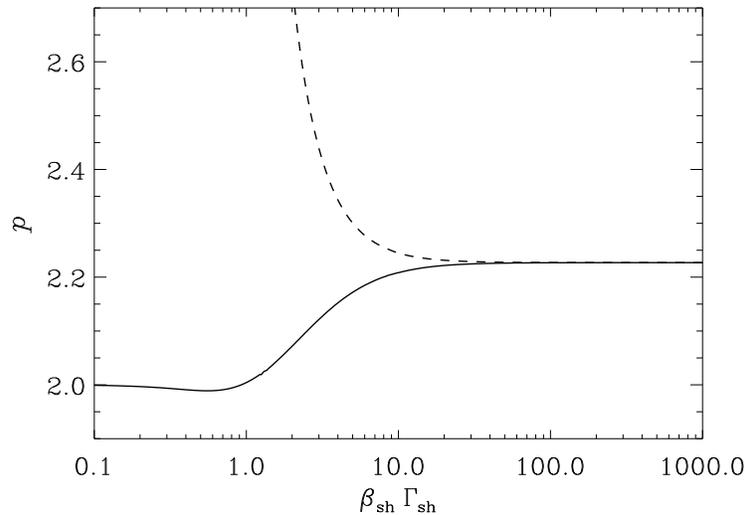}
\end{center}
\caption{\label{pindex}
Spectral index of the Fermi-accelerated particle
distribution as a function of the shock four-velocity,
for the two extreme cases shown in Fig.~\ref{velrat},
namely a cold upstream gas ({\it solid line\/}) and
an ultra-relativistically hot one ({\it dashed line\/})
}
\end{figure}

   For moderately relativistic shocks, the
self-consistent shock crossing angle distribution
and the spectral index depend on the shock jump
conditions assumed, as well as the shock Lorentz
factor $\Gammash$, which together determine the shock
velocity ratio.  Figure~\ref{pindex} shows, for
illustration, the spectral indices obtained with
the eigenfunction method of Kirk et al.\ \cite{Kiretc00}
in the two extreme cases considered in Sect.\
\ref{MRshocks}, namely that where the upstream gas
is `cold' so that the shock is strong, and that
where it is ultra-relativistically hot. 

   It is readily seen that while for high Lorentz
factors ($\Gammash \gesim 10$) the values obtained
rapidly converge to the ultra-relativistic case,
for lower shock Lorentz factors the different shock
jump conditions yield very different spectral indices.
In particular, for moderately relativistic shocks in
a relativistically hot gas, which might be relevant to
internal shocks in gamma-ray bursts, the spectral
indices obtained can be significantly steeper than
in the ultra-relativistic case.
For strong shocks, on the other hand, the spectral index goes
smoothly from the ultra-relativistic value of $p = 2.23$
to the non-relativistic one of $p = 2$ as the shock
Lorentz factor decreases.

   The above results apply to unmagnetised shocks
and isotropic direction-angle diffusion both upstream
and downstream.  A non-negligible magnetisation parameter
(\ref{sigma}) lowers the strong shock velocity
ratio, leading to steeper values of the spectral index
than in the corresponding unmagnetised case \cite{Kiretc00}.
Anisotropic diffusion in direction angle can have the
opposite effect: calculations for the extreme case where the
magnetic fluctuations are concentrated in the plane of
the shock yield somewhat flatter spectra than the
isotropic case, but by less than 0.1 in the spectral
index $p$ \cite{Kiretc00}.  Anisotropic \emph{pitch-angle}
diffusion has been simulated by Bednarz and Ostrowski
\cite{BedOst98}, who obtained steeper spectral indices
for weak scattering, the regime considered above
corresponding to the limit of strong scattering.

\section{Maximum and Minimum Particle Energies}
\label{maxmin}
   I now turn to the question of the range in particle
energies over which the spectrum derived above
applies.  I first discuss the acceleration time scale
for the Fermi mechanism, then use it to derive the
maximum energy attainable by this process at a
relativistic blast wave, and consider an alternative
scenario involving the initial boost which can reach
higher energies.  
I also discuss the minimum energy for Fermi acceleration,
and the need for a distinct electron pre-acceleration
mechanism in electron--ion shocks.

\subsection{Acceleration Time Scale}
   The acceleration time scale $t_\mathrm{acc}$ is
defined as the time needed for the particle energy
to increase by an amount of order itself.  Since,
as seen in Sect.\ \ref{E-updyn}, this typically
occurs every shock crossing cycle, $t_\mathrm{acc}$
is roughly the cycle time, which is the sum of
the upstream and downstream \emph{residence
times} $t_\mathrm{up}$ and $t_\mathrm{dn}$.
In the case of deflection by a uniform magnetic
field upstream, the former is of order
\beq
\label{tup-defl}
t_\mathrm{up} \sim \frac{1}{\Gammash \omega_{\mathrm{c}\perp}}
\equiv \frac{E}{q \Gammash B_{1\perp} c} \; ,
\eeq
where $q$ and $\omega_\mathrm{c}$ are the
particle's charge and cyclotron frequency.
   The downstream residence time depends on the
downstream scattering process; assuming Bohm
diffusion, it is roughly the downstream gyrotime,
\beq
t_\mathrm{dn}' \sim \frac{1}{\omega_\mathrm{c}'}
\equiv \frac{E'}{q B_2' c} \; ,
\eeq
where primed quantities are measured in the downstream
rest frame, as before.  If the downstream magnetic field
is simply the compressed value resulting from the (weakly
magnetised) ultra-relativistic shock jump conditions, $B_2'
\approx B_{2\perp}' \approx \sqrt{8} \Gammash B_{1\perp}$,
it can be shown that $t_\mathrm{dn} \sim t_\mathrm{up}$
\cite{GalAch99}.

   Turbulence downstream may amplify the magnetic field
$B_2'$ by a significant factor above the shock-compressed
value, thereby reducing the downstream residence
time by the same factor; assuming the field reaches
equipartition with the thermal pressure downstream,
this factor will be of order $c/v_\mathrm{A}$, where
$v_\mathrm{A}$ is the upstream Alfv\'en speed.
On the other hand, in the case of scattering by
small-scale magnetic fluctuations, the upstream
residence time is increased from the value
(\ref{tup-defl}) by a factor of order $R_\mathrm{L} /
(\Gammash \ell)$, where $\ell$, as before, is the
correlation length of the magnetic field \cite{GalAch99}.
Thus the value of $t_\mathrm{up}$ given in (\ref{tup-defl})
is a lower limit to $t_\mathrm{acc}$ in all cases of
interest.

\subsection{Relativistic Blast Waves and Ultra-High-Energy
            Cosmic Rays}
   An immediate application of the above
considerations is to the maximum energy attainable by
Fermi acceleration at the relativistic blast waves
occurring in fireball models of gamma-ray bursts; I thus
first review some basic properties of these models.
After an acceleration stage, an initially radiation-dominated
fireball enters a relativistic `free expansion' phase, in
which a blast wave is driven into the surrounding medium
with the approximately constant Lorentz factor $\Gammash
\approx \sqrt{2} \eta$, where $\eta \equiv \mathcal{E} /
(M c^2)$, $M$ being the baryonic mass in which the fireball
energy $\mathcal{E}$ is initially deposited \cite{MesLagRee93,%
PirSheNar93}.  This is followed by an adiabatic deceleration
phase in which the blast wave Lorentz factor decreases with
radius $R_\mathrm{sh}$ as $\Gammash \propto R_\mathrm{sh}^{-3/2}$
\cite{BlaMcK76}.  The transition between these two phases
occurs around the deceleration radius $R_\mathrm{dec}$,
given by
\beq
R_\mathrm{dec} \approx \left( \frac{3}{4\pi} \frac{\mathcal{E}}%
                       {\eta^2 \varepsilon_1} \right) \; ,
\eeq
where $\varepsilon_1 \approx n_1 m c^2$ is the energy
density of the surrounding material.

   In the absence of energy loss processes, the maximum particle
energy attainable by Fermi acceleration is set by the requirement
that $t_\mathrm{acc}$ be shorter than the age of the system, which
for a relativistic blast wave is simply $R_\mathrm{sh} / c$.
Using (\ref{tup-defl}) for $t_\mathrm{acc}$, the resulting maximum
energy at a given blast wave radius $R_\mathrm{sh}$ is
\beq
\label{Emax-Rsh}
E_\mathrm{max} \approx q B_1 \Gammash R_\mathrm{sh} \; .
\eeq
Note that this is larger by a factor $\Gammash$ than a commonly
used estimate resulting from a simple geometrical comparison
of the particle Larmor radius with $R_\mathrm{sh}$ \cite{Hil84}.
This is due to features specific to particle acceleration at
a relativistic blast wave, in particular the fact that an
accelerated particle typically executes
only a fraction $\sim 1/\Gammash$ of a Larmor orbit upstream
before recrossing the shock.

   The evolution of the product $\Gammash R_\mathrm{sh}$ with
$R_\mathrm{sh}$ implies that the highest $E_\mathrm{max}$
is reached at the deceleration radius, $R_\mathrm{sh}
\approx R_\mathrm{dec}$.  It has the numerical value
\beq
\label{Emax}
E_\mathrm{max} \approx 5 \times 10^{15} \; Z B_{-6}
   \left( \frac{\mathcal{E}_{52} \eta_3}{n_0} \right)^{1/3}
   \; \mathrm{eV} \; ,
\eeq
for particles of charge $q = Z e$, where $B_{-6}$ is the
upstream magnetic field, $\mathcal{E}_{52}$ the (isotropic)
fireball energy, $\eta_3$ the initial Lorentz factor and
$n_0$ the upstream density, respectively in units of
microgauss, $10^{52}$\,erg, $10^3$ and $\mathrm{cm}^{-3}$,
these normalising values being those appropriate for a
GRB fireball expanding into a generic interstellar medium
\cite{GalAch99}.

   Equation (\ref{Emax}) rules out the production of
ultra-high-energy cosmic rays (UHECRs), with energies up
to $\sim 10^{20}$\,eV, at the \emph{unmodified, external}
blast waves of relativistic fireballs.  Scenarios which
postulate a GRB origin for UHECRs \cite{MilUso95,Vie95,Wax95}
must thus invoke some other site or mechanism to reach the
required particle energies.  The idea most often put forward
is that UHECRs are accelerated at internal shocks, where
substantially higher magnetic fields, close to equipartition,
could be present both upstream and downstream and allow the
Fermi mechanism to reach UHECR energies.  However, the
particle spectrum in this case would likely be steeper,
as argued above, reducing the efficiency of UHECR production.
Moreover, the important issue of the escape of these particles
from the interior of the fireball to the surrounding medium
remains to be investigated, as they could suffer significant
adiabatic losses due to the fireball expansion before escaping.

   Another possibility for acceleration at the blast wave
is that the upstream magnetic field
might be amplified by a large factor above its undisturbed
value due to instabilities driven by the accelerated particles
themselves, as was recently proposed in the context of
supernova remnants by Bell and Lucek \cite{BelLuc01}.
It is unclear, however, how far such instabilities would
have time to develop in the relativistic context, given the
comparatively short time before the modified upstream medium
is overtaken by the shock.

\subsection{Fireballs in Pulsar Wind Bubbles}
   An alternative scenario for UHECR acceleration
is based on the observation that
the \emph{initial boost} examined in Sect.~\ref{E-updyn} can
circumvent the age limit (\ref{Emax}), as it involves only
a downstream half-cycle.  The maximum energy is then
set only by the requirement that the downstream residence
time $t_\mathrm{dn}$ be less than the age of the system.
If one assumes that the downstream magnetic field is
turbulently amplified close to equipartition values,
energies of order $10^{20}$\,eV or more can be reached,
provided that a population of relativistic particles
with sufficient initial energy to be boosted into this
range is present upstream \cite{GalAch99}.

   Galactic cosmic rays with appropriate energies are present
in the interstellar medium, but constitute only a small fraction
of the upstream energy density, so that only a correspondingly small
fraction of the fireball energy could go to boost these to UHECR
energies.  However, a situation where the surrounding medium
consists almost exclusively of relativistic particles of the
required energy occurs naturally in the context of neutron
star binary merger events: the close binary pulsar systems
observed in our Galaxy, which are the progenitors of these
merger events, all contain millisecond pulsars with
characteristic spindown times of order
$10^8$\,yr, while their spiral-in times due to
gravitational radiation are of order $3 \times 10^7$\,yr
\cite{NarPacPir92}.  These pulsars thus fill the
surrounding space with relativistic particles over the
lifetime of the binary system, forming
a large pulsar wind bubble in the interstellar medium.

   While the majority constituents of these pulsar wind
bubbles will likely be electron-positron pairs, pulsar winds
also seem to contain ions \cite{GalAro94,Hosetc92}.  Scaling
this ion component to millisecond pulsar
parameters, one can show that it yields ions with energies
$\sim 10^{14}$\,eV, sufficient to be boosted to UHECR energies
provided $\Gammash \gesim 10^3$.  This process is now highly
efficient: a large fraction of the fireball energy can go
to boost these ions to UHECR energies.  Moreover, for typical
parameters the blast wave will decelerate within the pulsar
wind bubble, resulting in a power-law spectrum of boosted
ions,
\beq
\frac{\D N}{\D E} \propto E^{-2} \; ,
\eeq
with a lower bound of $\sim 3 \times 10^{18}$\,eV, compatible
with the inferred UHECR source spectrum \cite{GalAch99}.
This scenario thus naturally provides for the acceleration
of UHECRs into the required energy range, with the required
spectrum, and with high efficiency.

\subsection{Minimum Energy and Electron Pre-Acceleration}
\label{preacc}
   Returning now to the Fermi acceleration mechanism proper,
it has as one of its requirements that accelerated particles
see the shock as a sharp discontinuity, as its treatment in
Sect.~\ref{Fermi} makes clear.  For this to be the case,
the particle Larmor radius must be larger than the shock
thickness, which is in turn roughly given by the downstream
thermal ion Larmor radius.  Ions can thus undergo Fermi
acceleration when they have reached a few times their
downstream thermal energy, but electrons in electron--ion
shocks must first reach a minimum energy
\beq
\label{Emin}
E_\mathrm{min}' \sim \Gammarel m_\mathrm{i} c^2 \; ,
\eeq
where $m_\mathrm{i}$ is the ion mass, before participating
in the Fermi mechanism and acquiring its characteristic
spectral index.

   Unless one assumes that these objects involve solely
electron--positron shocks, the presence of synchrotron-emitting,
Fermi-accelerated electrons in GRB afterglows and Crab-like
supernova remnants thus requires an electron pre-ac\-ce\-ler\-at\-ion
mechanism.  This mechanism must bring the electron energy from that
resulting from randomisation of the bulk upstream energy, which as
seen in Sect.~\ref{MRshocks} is $E_\mathrm{th}' = \Gammarel m_\mathrm{e}
c^2$, to $E_\mathrm{min}'$, a factor of the mass ratio
$m_\mathrm{i} / m_\mathrm{e}$
higher. An acceleration process operating over precisely this energy
range is the resonant ion cyclotron wave absorption mechanism
discovered by Hoshino et al.\ \cite{Hosetc92} in numerical
simulations of highly relativistic, electron--positron--ion
shocks.

   This resonant ion cyclotron acceleration mechanism typically
yields harder power-law spectra than those resulting from
Fermi acceleration: the spectral indices $p$ obtained
from the simulations are generally less than 2, and a
value as low as $p=1$ is predicted in a quasi-linear,
steady-state approximation \cite{Hosetc92}.
The resulting picture for the accelerated electron (and
positron) spectrum in an ultra-relativistic shock containing
ions is thus of a relatively hard power-law spectrum at
low energies, steepening to the $p \approx 2.2$ spectrum
characteristic of Fermi acceleration at a break energy given
by (\ref{Emin}).  This might explain the flat radio spectral
indices of Crab-like supernova remnants, as well as the two breaks
in the Crab Nebula spectrum between radio and X-ray frequencies,
only one of which can be attributed to synchrotron cooling.

\section{Summary}
   The shock velocity ratio $r$ across a relativistic shock
is in general a function of the assumed upstream temperature
as well as the shock Lorentz factor $\Gammash$, but it rapidly
tends to the ultra-relativistic limit $r=3$ for $\Gammash \gesim
10$.  The ultra-relativistic Fermi acceleration regime then mirrors
some of the simplicity of the non-relativistic, strong shock regime,
this asymptotic shock velocity ratio corresponding to an asymptotic
power-law index of the accelerated particle distribution.
For the specific case of isotropic direction-angle
scattering on both sides of the shock, this spectral index
is $p = 2.23 \pm 0.01$; more generally, a value of $p$ in the
range 2.2--2.3 is found under a variety of particle
transport assumptions.  These values are consistent with the
observed spectra of sources thought to contain ultra-relativistic
shocks, such as gamma-ray burst afterglows and Crab-like
supernova remnants.  For moderately relativistic shocks,
the spectral index depends on the shock jump conditions as
well as $\Gammash$; in particular, shocks in a relativistic
gas typically yield steeper spectral indices than the above
ultra-relativistic values.

   The maximum energy $E_\mathrm{max}$ of the Fermi-accelerated
particle distribution is determined by the acceleration time,
which is in general set by the upstream residence time.  For
acceleration at the unmodified, external blast wave of relativistic
fireballs, this yields $E_\mathrm{max} \sim 10^{16}$\,eV
for typical parameters of the surrounding interstellar medium,
ruling out the production of ultra-high-energy cosmic rays in
this context.  If neutron star binary merger events give rise
to relativistic blast waves with $\Gammash \gesim 10^3$, these
can provide an alternative scenario for UHECR production: ions
accelerated in the pulsar wind present before the merger can be
boosted to energies $\gesim 10^{20}$\,eV by the blast wave with
high efficiency; deceleration of the blast wave in the pulsar
wind bubble yields a spectral index $p = 2$ and a typical lower
cutoff around $3 \times 10^{18}$\,eV.
There is also a minimum energy for Fermi acceleration, set
by the requirement that the shock thickness be small relative
to the particle Larmor radius.  In electron--ion shocks, this
requires a distinct pre-acceleration mechanism for the
electrons, which could be the resonant ion cyclotron wave
acceleration mechanism of Hoshino et al.\ \cite{Hosetc92}.

\section*{Acknowledgements}

   I gratefully acknowledge support from the Italian Ministry
of University and Research through grant Cofin--99--02--02,
and the Netherlands Organisation for Scientific Research (NWO)
through GBE/MPR grant 614--21--008.

%

\end{document}